\documentclass[sn-mathphys,Numbered]{sn-jnl}% Math and Physical Sciences Reference Style
\usepackage{graphicx}%
\usepackage{multirow}%
\usepackage{amsmath,amssymb,amsfonts}%
\usepackage{amsthm}%
\usepackage{mathrsfs}%
\usepackage[title]{appendix}%
\usepackage{textcomp}%
\usepackage{manyfoot}%
\usepackage{booktabs}%
\usepackage{algorithm}%
\usepackage{algorithmicx}%
\usepackage{algpseudocode}%
\usepackage{listings}%
\usepackage{pdfpages}
\usepackage{doi}
\usepackage{enumitem}%itemize leftmargin

\usepackage{tikz}
\usetikzlibrary{arrows.meta}  % Für bessere Pfeilspitzen

\begin{document}

\title{Extending the European Competence Framework for Quantum Technologies: new proficiency triangle and qualification profiles}

\author*[1]{\fnm{Franziska} \sur{Greinert}}\email{f.greinert@tu-braunschweig.de}

\author[2,3]{\fnm{Simon} \sur{Goorney}}

\author[1]{\fnm{Dagmar} \sur{Hilfert-Rüppell}}%

\author[1]{\fnm{Malte S.} \sur{Ubben}}%\email{malte.ubben@tu-braunschweig.de}

\author[1]{\fnm{Rainer} \sur{Müller}}%\email{rainer.mueller@tu-braunschweig.de}

\affil[1]{\orgdiv{Institut für Fachdidaktik der Naturwissenschaften}, \orgname{Technische Universität Braunschweig}, \orgaddress{\street{Bienroder Weg 82}, \city{Braunschweig}, \postcode{38106}, \country{Germany}}}

\affil[2]{\orgdiv{Department of Management}, \orgname{Aarhus University}, \orgaddress{\street{Fuglesangs Allé 4}, \city{Aarhus}, \postcode{8000}, \country{Denmark}}}
\affil[3]{\orgdiv{Niels Bohr Institute}, \orgname{Copenhagen University}, \orgaddress{\street{Blegdamsvej 17}, \city{København}, \postcode{2100}, \country{Denmark}}}

\abstract{
With the increasing industrial relevance of quantum technologies (QTs), a new quantum workforce with special qualification will be needed. Building this workforce requires educational efforts, ranging from short term training to degree programs.
In order to plan, map and compare such efforts, personal qualifications or job requirements, standardization is necessary. The European Competence Framework for Quantum Technologies (CFQT) provides a common language for QT education.
The 2024 update to version~2.5 extends it by the new proficiency triangle and qualification profiles: The proficiency triangle proposes six proficiency levels for three proficiency areas, specifying knowledge and skills for each level. Nine qualification profiles show prototypical qualifications or job roles relevant to the quantum industry, with the required proficiency, examples, and suggestions. This is an important step towards the standardization of QT education.
The CFQT update is based on the results of an analysis of 34 interviews on industry needs. 
The initial findings from the interviews were complemented by iterative refinement and expert consultation. 
}

%3-10
\maketitle

\newcommand{\cf}{CFQT}

\section{Introduction}\label{sec:Intro}
Modern Quantum Technologies (QTs), such as quantum computers, quantum sensors or quantum communication devices, are gaining increasing industrial relevance. Building the future quantum workforce requires not only intensive educational efforts such as training and study programs. 
QT education also needs a common language as well as standardization efforts to make education and qualifications comparable: The European Competence Framework for Quantum Technologies (\cf)~\cite{greinertEuropeanCompetenceFramework2024a} is the ``reference framework for planning, mapping and comparing QT-related educational activities, personal qualification and job requirements''. This paper documents the update from version~2.0 to version~2.5 and the advances in the new version.

The \cf\ aims to structure QT competences as well as related competences relevant to the future quantum workforce. It was already used, for example, to map course content~\cite{hellsternIntroducingQuantumInformation2024}, to plan new courses~\cite{patinoQuantumComputingEducation2024}, to categorize research~\cite{engelsbergerQuantumreadyVectorQuantization2023}, or as the backbone for a curriculum transformation framework~\cite{goorneyFrameworkCurriculumTransformation2024}.
Version 2.5 (2024) of the \cf\ was published by the European Commission, EU %Directorate-General for Communications Networks, Content and Technology, 
Publications Office (see Ref.~\cite{greinertEuropeanCompetenceFramework2024a}). % of the European Union. 
It was developed within the European Quantum Flagship~\cite{qucatsQuantumFlagshipFuture2024} Coordination and Support Action (CSA) projects QTEdu~\cite{qteducsaQuantumTechnologyEducation} and QUCATS~\cite{europeancommissionQuantumFlagshipCoordination2024}. 
Within QUCATS, the European Quantum Readiness Center (EQRC)~\cite{quantumflagshipEuropeanQuantumReadiness}
has been launched to establish best practices, including standardization, using the \cf\ as a tool. 

The objective of the update to version~2.5 was to provide more detailed and substantiated descriptions for the proficiency levels, as well as a new version for the  qualification profiles from 2022~\cite[beta version,][]{greinertQualificationProfilesQuantum2022}. General descriptions for six proficiency levels were  already added in the \cf\ update to version~2.0, but needed further refinement and specification for QT -- as in the new \textit{proficiency triangle} (see Sec.~\ref{sec:PT}). 
The proficiency triangle enables the specification of proficiency or qualification in three dimensions, covering (I)~quantum concepts, thus physics fundamentals, (II)~QT engineering competences, including QT functionalities, and (III)~QT applications and strategies,  i.e. the business perspective, including aspects such as impact and ethics. 
By assigning a proficiency level for each of these three proficiency areas, the qualification of an individual can be specified, for example.  

%Based on the interview analysis, a 
A total of nine new \textit{qualification profiles} were identified and visualized with the proficiency triangle (see Sec.~\ref{sec:QP}). They represent prototypical qualifications relevant to industry. 
However, to really specify such a qualification, a second measure is needed: A selection from the \textit{content map}. 
It structures content relevant in the context of QT and was in the focus of the update to version~2.0~\cite{greinertQuantumReadyWorkforce2023a}. %, structured in the \textit{content map}. 
The content map and the details for the eight content domains remained consistent from version~2.0 (2023) to 2.5~(2024). In the new version, two examples  are given of how to combine such a selection from the content map and a profile based on the proficiency triangle. %With the update, two key additions have been made: the \textit{proficiency triangle} (see Sec.~\ref{sec:PT}) and the \textit{qualification profiles} (see Sec.~\ref{sec:QP}). 
Appendix~\ref{app:structure} provides an overview of the general structure of the \cf\ document, from the content map and the proficiency triangle to the qualification profiles and examples. 

Crucial input for the \cf\ extension in version~2.5 came from the analysis of 34~interviews focusing on QT educational needs in industry. 
The primary objective of the interview analysis was to report on the qualification and training needs in industry, as documented in %\textit{Advancing quantum technology workforce: industry insights into qualification and training needs}
Ref.~\cite{greinertAdvancingQuantumTechnology2024}. However, the insights from the interviews provided great input for the \cf\ update: key competences were identified and incorporated into the proficiency level descriptions, and the jobs and roles discussed in the interviews led to the first draft of the new qualification profiles. 
In addition to the manual analysis, a GPT was employed to analyze anonymized interview transcripts and extract typical job roles
to improve the initial qualification profiles.

\section{Methods and procedure for updating to version 2.5}
%\subsection{Initial development and interview analysis}\label{subsec:methods:initial_interviews}
The development of the \cf\ started with an iterative study in 2020/2021~\cite{greinertFutureQuantumWorkforce2023}, leading to version~1.0~\cite{greinertCompetenceFrameworkQuantum2021a} compiled within the QTEdu CSA~\cite{qteducsaQuantumTechnologyEducation}. Within the QUCATS project~\cite{europeancommissionQuantumFlagshipCoordination2024}, it was updated to version~2.0, as documented in Ref.~\cite{greinertQuantumReadyWorkforce2023a}.
In this update, the content map was revised and descriptions of six proficiency levels A1 to C2 were added. The proficiency level descriptions were based on the European Qualifications Framework (EQF)~\cite{europeancommissionEuropeanQualificationsFramework2018}, thus specifying the required knowledge and skills for each proficiency level adapted to QT in general.

The proficiency level labels~A1, A2, B1, B2, C1 and~C2 are associated with the levels commonly used to specify qualifications in languages in Europe~\cite{councilofeuropeCommonEuropeanFramework2020a}. The DigCompEdu~\cite{redeckerEuropeanFrameworkDigital2017a} framework was used as a template which also provided the initial version of the level keywords, e.g. awareness for level~A1. These keywords were also updated in version~2.5 to better match the level descriptions.

The update to version~2.5, which brought the proficiency triangle and qualification profiles into the framework, was released in April 2024. The process and results of the update are documented below, including some excerpts from the \cf\ document for illustration. A version history and related publications are given in Appendix~\ref{app:versHis}. %, Table~\ref{tab:versions}. 
%They were converted from British to American English for consistency. 

\subsection{About the interviews: introduction of `competence types'}\label{subsec:methods:compTypes}
In summer~2023, 34~interviews with industry representatives were conducted and analyzed to answer research questions including
\textit{What QT qualification and training needs are reported by industry?} An in-depth analysis is documented in Ref.~\cite{greinertAdvancingQuantumTechnology2024}, including details about the methodology, which was based on qualitative content analysis~\cite{mayringQualitativeContentAnalysis2014}. 
The interviewees represent QT companies or companies with QT departments from different European countries, from start-ups to very large companies, covering technical, management or business roles.

In some of the 
interviews, a slide about the `competence types' (reproduced in 
Fig.~\ref{fig:CompTypes}) was shown to stimulate the flow of the interviews -- others were conducted without this input, e.g. if the interviewee expressed  very clear or different views in the previous questions. 
 %Three levels were listed for each of the three `competence types': build/develop QT (level \mbox{B--Build...}, \mbox{C--Component} improvement..., D--Develop...), overview/communicate about QT (level %N, O, P 
% N--Notice..., O--Overview..., P--Perform comparison...
% ), and use/adapt QT (%U, V, W%
% \mbox{U--Use/}run..., \mbox{V--Variate/}adapt..., W--Wrap...together...
% ), see Appendix~\ref{app:material_CT}. 
%
%\vbox{%
%\noindent
%The three competence types shown in Fig.~\ref{fig:CompTypes} with three levels each are:
This material introduced a structure of competence descriptions for three levels for each of the three `types' \textit{overview, communicate} (level N, O, P), \textit{build, develop} (level B, C, D) and \textit{use, adapt} (level U, V, W):
\vspace{.5\baselineskip}

\noindent
\hspace{0.05\textwidth}\begin{minipage}{0.9\textwidth}
\small     \itshape      \linespread{0.9}\selectfont
\textbf{Overview, communicate }(overview on applications and use cases)
\begin{itemize}
    \item[N] Notice the basic idea and potential of an application and (possible) use cases.
    \item[O] Overview of an application's landscape and use cases, critical perspective on potential and limitations, know and communicate with basic quantum vocabulary.
    \item[P] Perform a comparison of different applications based on a deeper understanding of the various functions, assess which application is suitable for which use case.
\end{itemize}
\end{minipage}

%}

\vspace{\baselineskip}
\noindent
\hfill\begin{minipage}[t]{0.49\textwidth}
\small     \itshape      \linespread{0.9}\selectfont 
\textbf{Build, develop} (build application)
\begin{itemize}
    \item[B] Build an application component, assemble.
    \item[C] Component improvement for an application.
    \item[D] Develop new components or applications through research, selection and integration of components.
\end{itemize}
\end{minipage}%
\hfill
\begin{minipage}[t]{0.49\textwidth}
\small     \itshape      \linespread{0.9}\selectfont
\textbf{Use, adapt} (use an application)
\begin{itemize}
    \item[U] Use/run an application for one specific use case that is customized for that use case.
    \item[V] Variate/adapt an application for several related use cases.
    \item[W] Wrap applications and components together to enable new use cases.
\end{itemize}
\end{minipage}\hfill
\vspace{\baselineskip}

\noindent
This was an outline for structuring typical competences that we expected to be relevant in industry. Thus, the key competences and qualifications discussed in the interviews were categorized according to this structure, regardless of whether the material was used in the interview or not. % of this material. 

\begin{figure}[b]
    \centering
    \includegraphics[width=1\linewidth]{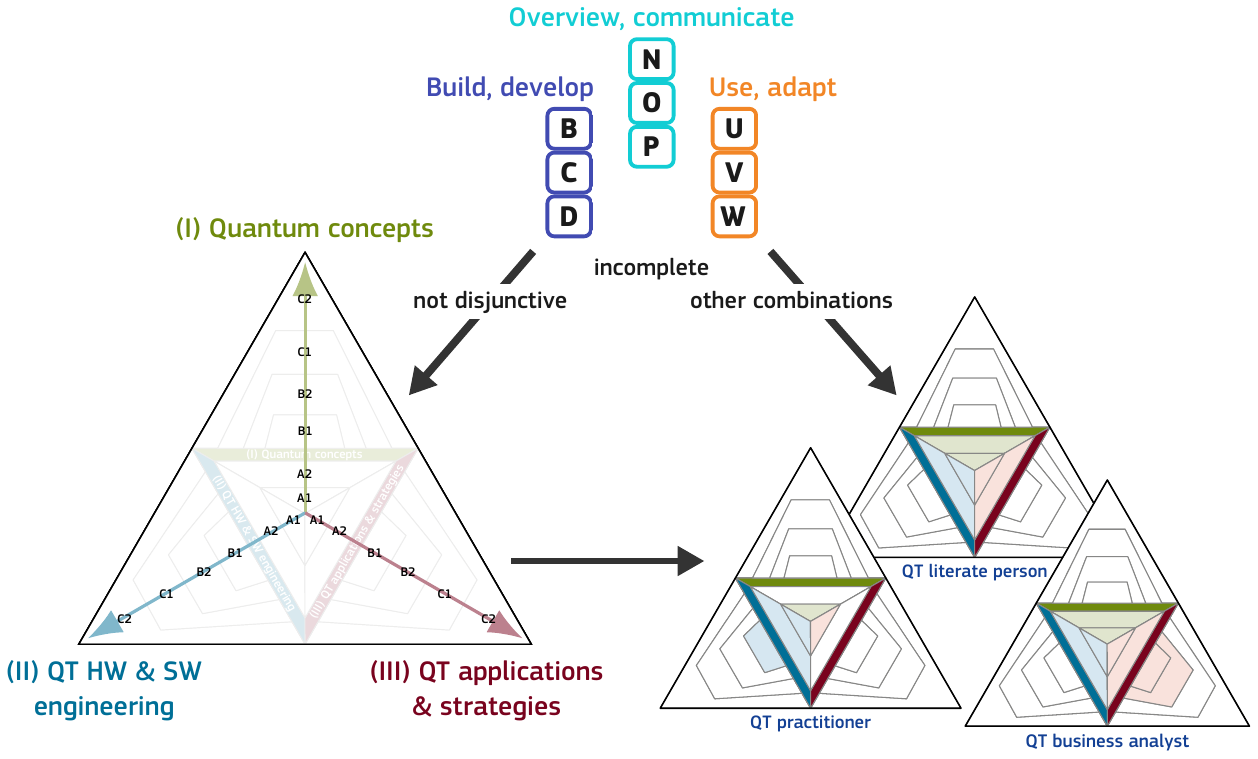}
    \caption{From the `competence types' (top) to the proficiency triangle (left) and the qualification profiles (right), which use the proficiency triangle as visualization. The issues of the `competence types' of being incomplete and not disjunctive descriptions of competences are addressed through the proficiency triangle. The second issue that rather combinations of competency type levels were discussed in industry is addressed through the qualification profiles.}
    \label{fig:fromCFtoPTQP}
\end{figure}

It turned out that this structure was incomplete and not disjunctive, so  refinement was needed. More concretely, the structure of the competence types was the starting point for developing or extracting both, the proficiency triangle and the qualification profiles, as discussed in the next sections and visualized in Fig.~\ref{fig:fromCFtoPTQP}: 
The proficiency triangle addresses the issue of disjunctive proficiency areas with independent proficiency levels for each area. In contrast, the qualification profiles show typical combinations of these levels, visualized as coverage of the proficiency triangle, for a more complete picture than the `competence types' provided.

\subsection{From `competence types' to the proficiency triangle}\label{subsec:methods:ProfTri}
Key competences were identified in the interview transcripts and analyzed not only in terms of the corresponding proficiency level, but also in terms of the three categories, which were initially the three `competence types' that iteratively evolved into the three proficiency areas: 
\begin{itemize}
    \item from ``overview [...] communicate with basic quantum vocabulary'', plus additions to area \textit{(I)~quantum concepts},
    \item  from `build, develop' and `use, adapt' to \textit{(II)~QT hardware (HW) \& software (SW) engineering}, and
    \item   aspects from `overview, communicate', e.g. ``critical perspective on potential and limitations'', plus additions to \textit{(III)~QT applications \& strategies}.
\end{itemize}%
Through another iterative process, the graphical appearance of a triangle was constructed.
In this way, the initial structure of the `competence types' was refined into what we named the \textit{proficiency triangle}, see Sec.~\ref{sec:PT}. 
It consists of three proficiency areas with six proficiency levels each. %, making a total of 18~levels. 

For each proficiency level,
a competence statement was formulated using action verbs%, with advice from experts in competence framework development
~\cite{greenECampusOntarioOpenCompetency2021}. 
They are complemented by descriptions of knowledge and skills based on the EQF levels~\cite{europeancommissionEuropeanQualificationsFramework2018}, as were the general level descriptions in version~2.0~\cite{greinertEuropeanCompetenceFramework2023,greinertQuantumReadyWorkforce2023a}. 
In contrast to the descriptions in version~2.0, the proficiency levels are now specified for the three proficiency areas and are based on the key competences identified through the interview analysis. Further details, such as relations to the content map, are also provided. All details for one proficiency level are given as an example in Sec.~\ref{sec:PT}.

%All proficiency levels were formulated in a generic way, adapted from the EQF definitions for QT in general.  
%In the new version~2.5, they are specified for the three proficiency areas, and key competences identified in the interviews were incorporated.  %The level keywords were also updated to better match the level descriptions.

\subsection{From the identification of job roles to qualification profiles}\label{subsec:methods:jobRoles_QualiProf}
%In addition, categories related to job roles were formulated. 
During the qualitative content analysis of the interview transcripts, twelve categories related to job roles were used. These categories were mainly based on the structure of the `competence types' (Sec.~\ref{subsec:methods:compTypes}/Fig.~\ref{fig:CompTypes}), as well as some frequently discussed combinations of two `competence type' levels as additional categories. 

In %Table~\ref{tab:categories}
Appendix~\ref{app:categories_QP}, these job role categories are shown, together with references to related `competence type' levels. For example, the  category ``engineer with overview on QT [O/B]'' is related to the ``O -- Overview...'' and ``B -- Build...'' and evolved to profile P4 of the QT practitioner, including also the ``advanced user [V]'', related to ``V -- Variate...''. Iteratively, the categories from the interview analysis evolved to the qualification profiles.

To support this process, 30 anonymized interview transcripts were analyzed using a large language model (LLM, namely GPT3.5%gpt3.5-turbo-16k
) to identify frequently mentioned job roles representing categories or clusters. This resulted in the descriptions of the 10~roles provided in Appendix~\ref{app:GPTpersonas}.
These roles were checked against the initial job role categories, in addition to intensive discussion of the initial categories. %, supported %the reduction from eleven to nine profile-related categories and an initial 
%a refinement of the job role categories. For these %remaining 
On this basis, the job role categories were refined, and a second iteration of categorizing the interview passages was performed. Several categories in the second iteration are already named similarly to the final profiles, as documented in Table~\ref{tab:categories}. Based on these categories, or more precisely, the categorized interview passages, the qualification profile descriptions version~2.1 were created. 

These profiles were then refined through discussions with several colleagues and by gathering feedback from experts in both academia and industry. For example, the profile names were changed to better represent industry roles, and the required proficiency levels for the profiles were discussed and adjusted based on these discussions. In addition to invited discussions with individual experts, an open event was conducted in March 2024 to gather feedback from the QTEdu  community~\cite[subpage Events]{qteducsaQuantumTechnologyEducation}. 

Table~\ref{tab:titleVersions} shows the evolution of the profile titles from version 2.1 to the published version. %In this way, t
The titles and descriptions were refined, and example personas/job roles and suggestions were added based on the results of the interview analysis documented in Ref~\cite{greinertAdvancingQuantumTechnology2024}. 
\begin{table}[ht]
    \centering
    \caption[Evolution of the titles of the qualification profiles.]{Evolution of titles of the qualification profiles. Although the titles may not have changed between two iterations, the descriptions may have changed significantly. These descriptions of the profiles in versions~2.1, 2.2, and 2.3 are available in the supplementary material on Zenodo~\cite{greinertSupplementaryMaterialAdvancing2024}.}
    \phantomsection\label{tab:titleVersions}
    \begin{tabular}{c|p{1.9cm}|p{2cm}|p{2.6cm}|p{3.7cm}}
        No. & \multicolumn{1}{c|}{v2.1} & \multicolumn{1}{c|}{v2.2} & \multicolumn{1}{c|}{v2.3} & \multicolumn{1}{c}{v2.5 (published in Ref~\cite{greinertEuropeanCompetenceFramework2024a})} \\
        \hline
        1 & \multicolumn{4}{c}{QT aware person} \\
        2 & \multicolumn{3}{c|}{QT aware decision maker} & QT informed decision maker\\
        3 & QT-literate communicator & \multicolumn{2}{c|}{QT communicator} & QT literate person\\
        4 & \multicolumn{2}{c|}{QT (market) analyst} & QT business analyst & \textbf{renumbered to Profile 5} \\
        5 & \multicolumn{2}{c|}{QT user, engineer working with QT} & QT technician & \textbf{Profile 4} QT practitioner\\
        6 & \multicolumn{2}{c|}{QT engineer (generalist)} & QT engineer/scientist (generalist) & QT engineering professional \\
        7 & \multicolumn{3}{c|}{senior QT engineer/QT architect} & QT (HW or SW) specialist\\
        8 & \multicolumn{3}{c|}{QT strategist} & QT (product) strategist \\
        9 & \multicolumn{4}{c}{QT core innovator}\\
    \end{tabular}
\end{table}

\subsection{Limitations}
The update of the \cf\ is primarily  based on the analysis of 34~interviews with participants from the quantum industry. This is a strength of the dataset, as it is directly constructed from interviews with experts in the field. However, we should anticipate and acknowledge that there may be  a bias  due to the group of interviewees, as discussed in Ref.~\cite{greinertAdvancingQuantumTechnology2024}. Most of the perspectives included come from people who have a very positive perception of QT, both in the interviews and in the iterative refinement. Therefore, they may identify QT-related job roles more readily than those without direct involvement in the industry would. 

The extraction of job roles from the interview transcripts by an LLM improves the objectivity of the qualitative content analysis by having a kind of second coder analyze most of the interviews (the 30 transcripts that could be anonymized) and prepare a second set of categories.  %: they were used -- together with expert discussions -- to refine the categories listed in Table~\ref{tab:categories} and thus influenced the data basis for the development of the qualification profiles.
However, the experts involved in the refinement of the qualification profiles must be expected to have a similar bias as the interviewees, since they have a similar background and experience.

In addition, the QT field is very active and rapidly changing. As the interviews were conducted in 2023, the identified needs that influenced the \cf\ update may also change in a short period of time. 
To capture these dynamic developments, annual updates of the \cf\ are planned as part of the QUCATS project, with the next one due in April 2025, and feedback is taken continuously.

%\section{Results: Proficiency triangle, qualification profiles and implications from the job post mapping}

\section{Proficiency triangle}%: 3 proficiency areas with 6 levels each}
\label{sec:PT}
The new proficiency triangle adds a second dimension, proficiency, to the content map from the \cf\ version~2.0. It consists of three proficiency areas (I)~quantum concepts, (II)~QT %hardware and software (
HW \& SW engineering, and (III)~QT applications \& strategies. 
The proficiency areas are described in the \cf\ document, for example:

\vspace{0.5\baselineskip}

\noindent\hspace{0.05\textwidth}\begin{minipage}{0.9\textwidth}
\small     \itshape      \linespread{0.9}\selectfont
    Area \textbf{(III) QT applications \& strategies} addresses the business dimension of QT applications (5.6, 6.7, 7.4, 7.6)%\footnote{The numbers 5.6, 6.7, ... refer to the subdomains from the content map.}
    . It focuses on the question of ``how to generate value with QT'' (domain 8). This also includes considerations of impact and responsibility, extending to the exploration of novel applications and the design of new products utilizing QT. \hfill \cite[quoted from][p.~14]{greinertEuropeanCompetenceFramework2024a}
\end{minipage}%

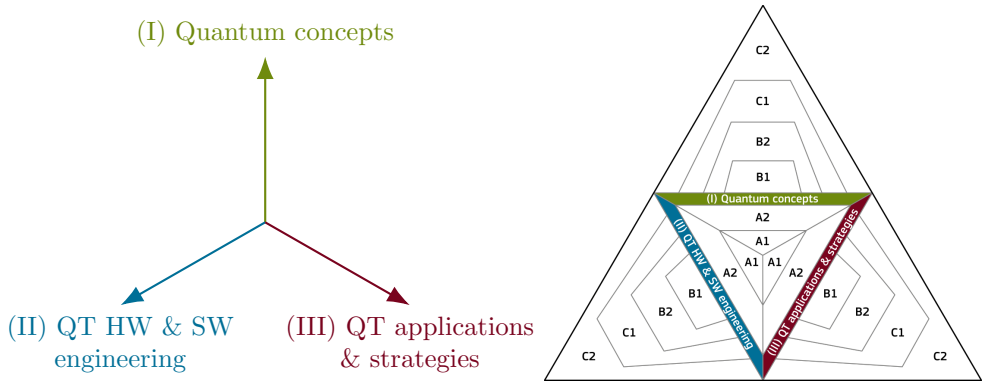
\begin{figure}[b]
    \centering
\begin{tikzpicture}
    \definecolor{area1}{RGB}{111,138,13}
    \definecolor{area2}{RGB}{0,111,151}
    \definecolor{area3}{RGB}{121,3,30}
    \def\axislength{2.2cm}  
    \draw[thick, area1, -{Latex[length=3mm]}] (0,0) -- (90:\axislength) node[above, area1] {(I) Quantum concepts};
    \draw[thick, area2, -{Latex[length=3mm]}] (0,0) -- (210:\axislength) node[below, area2, align=center] {(II) QT HW \& SW \\ engineering};
    \draw[thick, area3, -{Latex[length=3mm]}] (0,0) -- (-30:\axislength) node[below, area3, align=center] {(III) QT applications \\ \& strategies};
\end{tikzpicture}
\hspace{-3pt}\includegraphics[width=0.44\linewidth]{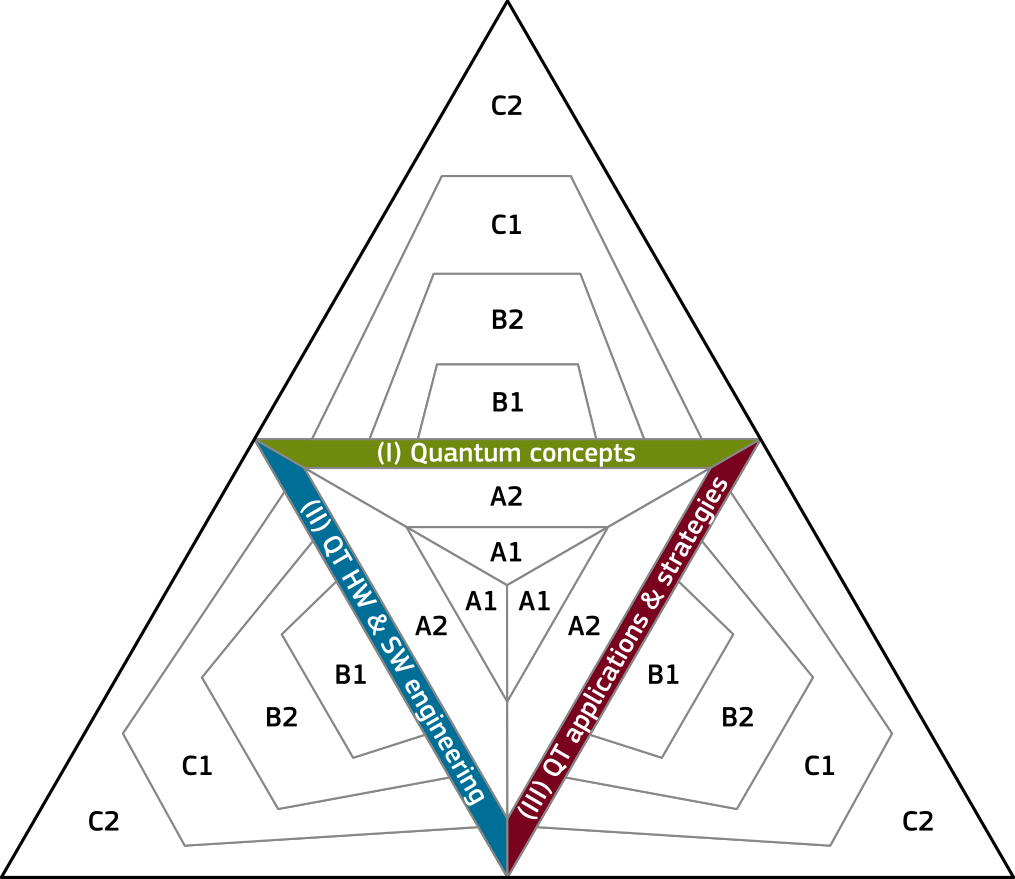}
    \caption[Proficiency areas and visualization in the proficiency triangle.]{The three proficiency areas (left) and their visualization in the proficiency triangle from the \cf~\cite[p.~30]{greinertEuropeanCompetenceFramework2024a} (right). The proficiency levels are labeled A1, A2, B1, B2, C1, and C2, with A~levels representing beginner, B~levels intermediate and C~levels advanced. 
    Proficiency first grows in breadth up to level~A2, preparing for specialization in the B~levels. At the tip of the triangle (level~C2), proficiency reaches and extends the state of the art, with strong specialization. However, proficiency also grows continuously in breadth, so that additional, e.g. rather B1~level, proficiency contributes to the higher levels, e.g. level~C2. 
    }
    \label{fig:profTriangleEmpty}
\end{figure}

\vspace{0.5\baselineskip}\noindent
The proficiency of an individual may grow in these three areas independently, however, typically someone working in the QT industry would have proficiency in all three areas. For each of these three proficiency areas, six proficiency levels~A1, A2, B1, B2, C1 and C2 are specified. Together they form the proficiency triangle shown in Fig.~\ref{fig:profTriangleEmpty}. With the coverage of the triangle, a QT-specific set of knowledge and skills, or a qualification can be visualized, as in the qualification profiles, % show prototypical coverage of the proficiency triangle:
%Each profile is visualized by a partially colored proficiency triangle, indicating the level of proficiency with respect to each of the three proficiency areas, 
see Sec.~\ref{sec:QP}.

%

%For each proficiency area, the six 
Each proficiency level consists of a competence statement, as well as knowledge and skills descriptions based on the EQF level definitions (as described in Sec.~\ref{subsec:methods:ProfTri}). The competence statements are reproduced in Table~\ref{tab:compStatem}. 
\begin{table}[t]
    \centering
    \caption[Competence statements (short version) from the \cf.]{Competence statements (short version) from the \cf~\cite[p.~14]{greinertEuropeanCompetenceFramework2024a}. 
    There are three additional pages in the \cf\ that provide more detail on the proficiency levels. A `QT facet' stands for a QT core, component, system, or application which can be HW and/or SW focused.%, `q.' for `quantum'.
    }
    \label{tab:compStatem}
    \begin{tabular}{p{0.195\textwidth-2\tabcolsep}
		|p{0.2673\textwidth-2\tabcolsep - 0.5\arrayrulewidth}
		p{0.2673\textwidth-2\tabcolsep - 0.5\arrayrulewidth}
		p{0.274\textwidth-2\tabcolsep}}
        Proficiency level & Area (I) & Area (II) & Area (III) \\
        \hline
        A1 Awareness & \textbf{Reproduce} basic quantum concepts \& terminology & \textbf{Reproduce} basic functionalities of a QT facet & \textbf{Recognize} potential of QT\\
        A2 Literacy & \textbf{Describe} fundamental quantum concepts & \textbf{Perform} basic tasks on a QT facet & \textbf{Identify} value of QT \\
        B1 Utilization & \textbf{Apply} quantum methods to problems & \textbf{Modify/apply} a QT facet & \textbf{Classify}~available~QT applications/approaches \\
        B2 Investigation & \textbf{Analyze} problems with quantum & \textbf{Analyze} performance, improve QT & \textbf{Analyze} QT market and opportunities\\
        C1 Specialization & \textbf{Refine} and extend quantum methods & \textbf{Conceptualize} integrated QT systems & \textbf{Advise} on QT appl. selection or strategies\\
        C2 Innovation & \textbf{Develop} innovative solutions & \textbf{Develop} new QT facet & \textbf{Develop} and assess QT (product) strategies\\
    \end{tabular}
\end{table}%
In addition, for each of the proficiency levels, 
relations to the content map (sub)domains (5.6, 6.7, ...) and examples of how the levels could be reached are given. 

For example, the description of the lowest proficiency level A1 for the proficiency area~(III)~QT applications \& strategies is \cite[quoted from][p.~17]{greinertEuropeanCompetenceFramework2024a}:
\vspace{0.5\baselineskip}

\noindent\hspace{0.05\textwidth}\begin{minipage}{0.9\textwidth}
\small     \itshape      \linespread{0.9}\selectfont
\textbf{A1 Awareness: Recognize potential of QT}
    \begin{itemize}\setlength{\itemsep}{0em}
        \item[\textbf{K:}] Basic idea of the potential of QT systems and applications, overview of possibilities, challenges and limitations.
        \item[\textbf{S:}] Ability to follow public media and discussions with critical awareness of hype.
    \end{itemize}
\end{minipage}
\vspace{0.5\baselineskip}

\noindent
For each level, related content (sub)domains \& examples are given, in this case:
\vspace{0.5\baselineskip}

\noindent\hspace{0.05\textwidth}\begin{minipage}{0.9\textwidth}
\small     \itshape      \linespread{0.9}\selectfont
    Applications for a QT, see 5.6, 6.7, 7.4, 7.6, or a selected (sub)topic (concrete application area, e.g. quantum optimization in logistics 5.6), with relevance for business~8.2 or education~8.4; hype (especially for computing).
\end{minipage}
\vspace{0.5\baselineskip}

\noindent
Examples of how to test or measure whether someone has reached a proficiency level are also provided. For this example, %level A1 in proficiency area (III) 
it is: \textit{Multiple choice questions on possibilities/limitations.} These remarks are a first step toward a certification scheme that will specify the measurement of proficiency in more detail.  

\begin{figure}[h]
    \centering
    \includegraphics[width=1\linewidth]{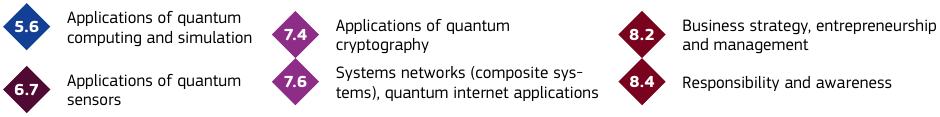}
    \caption{Related subdomains for proficiency level A1 in proficiency area (III), extracted from the content map~\cite[p.~5]{greinertEuropeanCompetenceFramework2024a}.
    }
    \label{fig:relSubd}
\end{figure}

The proficiency level descriptions are formulated independently of, e.g., a concrete QT pillar. To specify, e.g., a specific qualification of a person or the aim of a course or the requirements for a job, the proficiency triangle must be combined with a selection from the content map. For each level, there are references %The additions for each level refer 
to related content (sub)domains. In the example above, these are the subdomains 5.6, 6.7, 7.4, 7.6, 8.2 and 8.4. They are visualized in Fig.~\ref{fig:relSubd}.

All these subdomains of the content map influence the `flavor' of an individual's experience or the learning outcomes of a course, i.e. the specific technologies or concepts covered. A course may cover only applications in quantum computing (subdomain 5.6), or even only quantum optimization in logistics and thus only (sub)topics from subdomain 5.6 regarding applications. 
However, to cover level A1 in the business-related proficiency area (III), the relevance to business (8.2) needs to be covered with a critical perspective, e.g. addressing the hype in quantum computing. The focus would be on the value and impact of QT for an industry sector or the own company/business. 

Alternatively, the focus could be on the value and impact for society and education, i.e. subdomain~8.4 instead of~8.2. With 8.4 one can map qualifications outside of industry, e.g. a school teacher who recognizes the potential of QT as a context for teaching quantum concepts and who has an overview of how QT can impact society as a whole. This would be relevant, e.g., to discuss QT and their expected impact already in high school and thus to get more school students interested in quantum (and STEM). Raising this interest was rated as very necessary (40 of 52 participants rated `high need', 11 rated `low need', one rated `no need') in our follow-up survey to the interview analysis, see Ref.~\cite[p.~25%Sec.~5.4
]{greinertAdvancingQuantumTechnology2024}.

\section{Qualification profiles}\label{sec:QP}
The qualification of an individual or the objectives of a training, for example, can be visualized by a coverage of the proficiency triangle, which specifies the proficiency level for each of the three proficiency areas. Since these are formulated independently of the concrete subject matter, they have to be combined with a selection from the content map. 
Nine qualification profiles included in the \cf\ version~2.5 show the prototypical coverage of the proficiency triangle. They are listed in Table~\ref{tab:profiles} with the required proficiency levels for each of the three proficiency areas. 

\begin{figure}[t]
    \centering
    \includegraphics[width=0.9\linewidth]{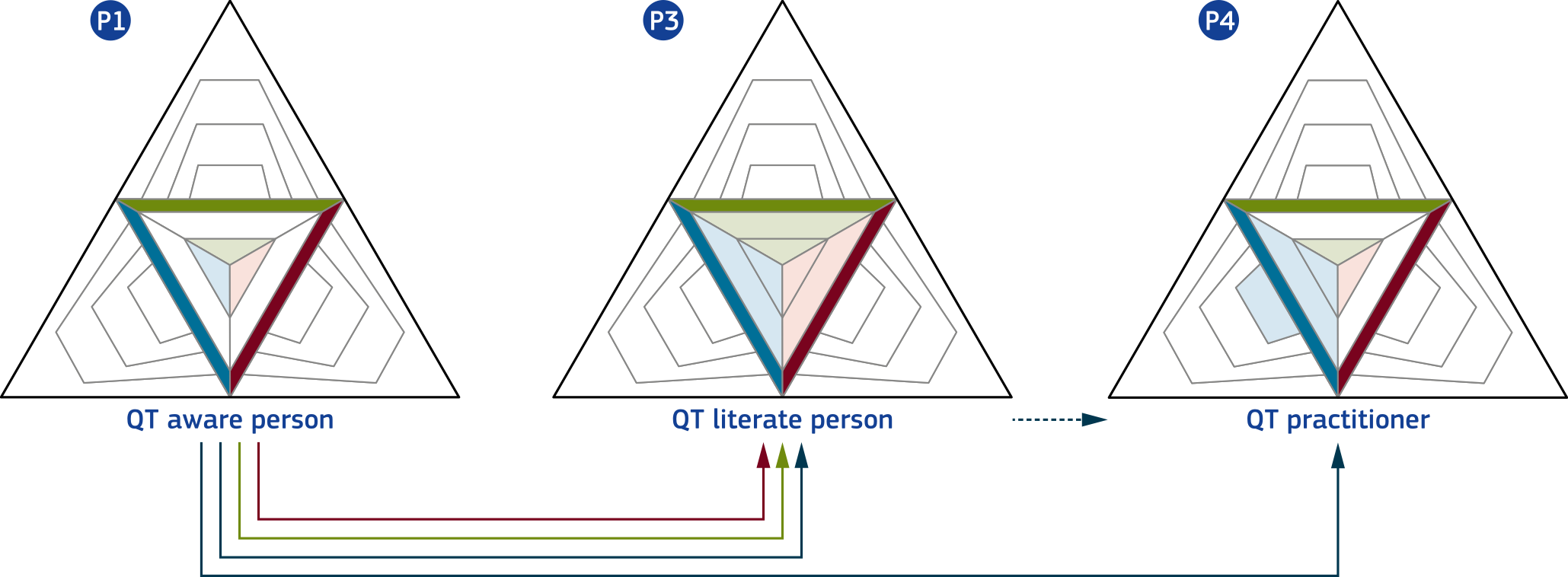}
    \caption[Three qualification profiles with profile relations from the CFQT.]{Three qualification profiles with profile relations, from the \cf\ summary~\cite[p.~30]{greinertEuropeanCompetenceFramework2024a}. The arrows visualize the profile relation, the solid line shows a typical progression, the dashed line shows another conceivable qualification path. The color of the lines corresponds to the color of the proficiency area in which the qualification grows from one profile to another.
    }
    \label{fig:threeProfiles}
\end{figure}

\begin{table}[b]
    \centering
    \caption{Qualification profiles and related proficiency levels for area (I) quantum concepts, (II) QT HW \& SW engineering and (III) QT applications and strategies 
    in the \cf\ version~2.5.}
    \begin{tabular}{cp{8.2cm}|ccc}
         &  & \multicolumn{3}{c}{Proficiency level in area} \\
        %& & ~~(I)~~ & ~(II)~ & (III) \\
        No. & Profile title& ~~(I)~~ & ~(II)~ & (III) \\%q. conc.&QT eng.&appl./st.\\
        \hline
        P1 & QT aware person & A1 & A1 & A1 \\
        P2 & QT informed decision maker & -- & -- &A1\\
        P3 & QT literate person (QT lit. business role, advocator, enthusiast)& A2&A2&A2\\
        P4 & QT practitioner (working with QT, technician, QT user) & A1 & B1 & A1 \\
        P5 & QT business analyst & A2 & A2 & B2\\
        P6 & QT engineering professional (e.g. QT engineer, quantum computer or information scientist) & B1 & B2 & A2\\
        P7 & QT (HW or SW) specialist (e.g. senior QT engineer, QT architect) & B2 & C2 & B1\\
        P8 & QT (product) strategist (e.g. advisor, business development expert) & B2 & B2 & C2\\
        P9 & QT core innovator & C2 & B1 & A2\\
    \end{tabular}
    \label{tab:profiles}
\end{table}
An overview page shows these nine profiles with the different, partly colored proficiency triangles as well as the relations between the profiles~\cite[p.~18]{greinertEuropeanCompetenceFramework2024a}.
Relations mean how people may evolve from one profile to another. As in the example in Fig.~\ref{fig:threeProfiles}, such a potential evolution is visualized by arrows.

For each profile, a detail page provides the partially colored proficiency triangle with the descriptions of the highest  proficiency level covered for each of the three proficiency areas. The profile is described generally, and example personas/job roles provide more concrete descriptions of what a job with that qualification might look like, for example for profile~4:
\vspace{0.5\baselineskip}

\noindent\hspace{0.05\textwidth}\begin{minipage}{0.9\textwidth}
\small     \itshape      \linespread{0.9}\selectfont
    \textbf{General description}
    The \textbf{QT practitioner} is someone who works around the development, assembly and operation of QT (technicians or ‘classical’ engineers with some QT  specific additional qualifications), or uses QT with some customization:
    \begin{itemize}
        \item is QT aware, thus is able to follow team discussions and has an idea of the potential of the QT working on,
        \item has an overview of the relevant parts (hardware and/or software) for QT, and
        \item focuses on the specific QT relevant to their own work, and knows how to work with it.
    \end{itemize}
    \textbf{Example personas} %[another one available in document~\cite{greinertEuropeanCompetenceFramework2024a})]\newline
    (‘Classical’ engineer with QT add-on, QT lab technician (e.g. for operation and maintenance), QT assembly and test technician, ...)
    \newline
    An \textbf{engineer working on QT development}, could be an electronic or mechanical engineer or a software engineer/computer scientist, working on the control hardware/software for a qubit, needs mainly traditional engineering skills, but works together with the quantum people, so needs an idea of the special challenges in QT development (but does not need to understand the details, has a supervisor who ensures compliance with quantum requirements) and also has an idea of the applications etc. to know what they are working for. [...] {\cite[p.~22]{greinertEuropeanCompetenceFramework2024a}}
\end{minipage}
\vspace{0.5\baselineskip}

\noindent
The profile description is complemented by a `needs and suggestions' section, describing the suggested previous qualification (i.e. another profile and/or a specific background), training modules to reach the profile starting from the suggested previous qualification. In addition, recommendations regarding language and certificates are provided. All these suggestions are based on the interview analysis. Some of the results documented in Ref.~\cite{greinertAdvancingQuantumTechnology2024} are included in these sections of the qualification profile detail pages. They include what training is considered relevant or what language is deemed appropriate.

\begin{figure}[ht]
    \centering
    \includegraphics[width=0.85\linewidth]{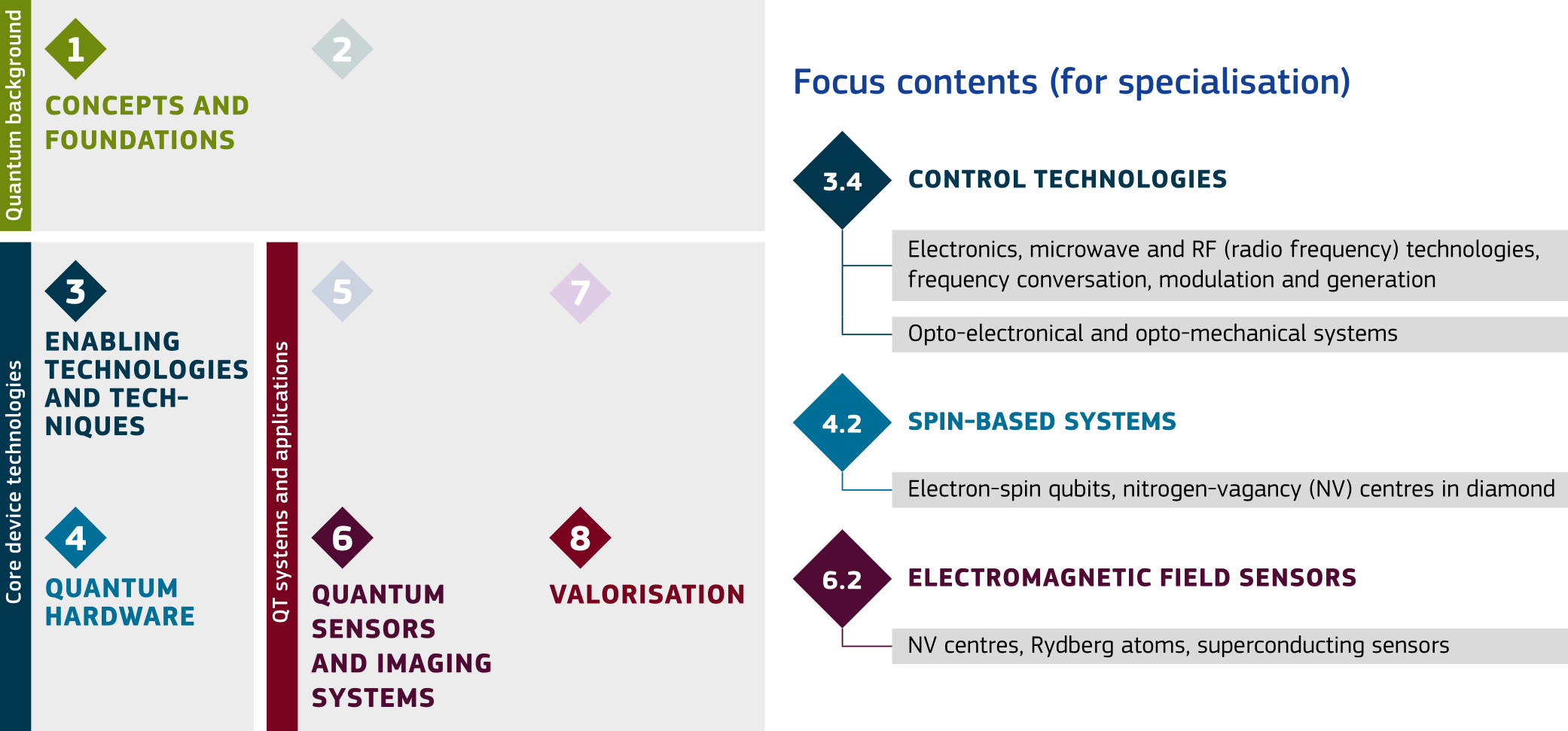}
    \caption[Content map coverage and focus topics (example from \cf).]{Content map coverage and focus topics for the \textit{quantum control electronics practitioner for NV sensors}~\cite[p.~28]{greinertEuropeanCompetenceFramework2024a}. On the left side, the covered domains from the content map are depicted: domains~1, 3, 4, 6 and 8. While the domains~1 and~8 are only covered at a low level, as indicated in the proficiency triangle (right triangle for P4 in Fig.~\ref{fig:threeProfiles}) with level A1 for proficiency area~(I) and~(III), the specialization is in area~(II) up to level~B1. This specialization is also shown on the right side (focus contents). Here, the selected subdomains with selected topics provide information about the concrete topics in the focus of the qualification. 
    } 
    \label{fig:QPexample}
\end{figure}

Finally it should be emphasized that the qualification profiles -- like the proficiency level description -- are formulated %a general framework 
independent of a specific technology in which they are applied. In order to describe a specific qualification, the qualification profile \textbf{--} or, more generally, the coverage of the proficiency triangle \textbf{--} must be combined with a topical selection from the content map. Two examples are given in the \cf: the \textit{quantum control electronics practitioner for NV sensors} and the \textit{quantum optimization in logistics \& production analyst}~\cite[p.~28~\&~29]{greinertEuropeanCompetenceFramework2024a}.
Figure~\ref{fig:QPexample} shows the coverage of the content map along with the focus topics for the first of these examples. The clear focus on quantum sensing and control technologies is visible. They complement the partially colored proficiency triangle for the QT practitioner (P4, right triangle in Fig.~\ref{fig:threeProfiles}): it shows the coverage of proficiency level~A1 for the proficiency areas~(I) and (III) along with level B1 in area (II). Each of the two examples is completed by a description of the example and the associated QT specific qualification.

\section{Discussion and conclusion}\label{sec:disConv25}
Prior to implementing new courses and programs to develop the workforce pipeline in QT, it is critical to answer a key question: What specific knowledge and skills are needed for industry growth? Different studies were conducted to analyze these needs~\cite{foxPreparingQuantumRevolution2020, aielloAchievingQuantumSmart2021,hughesAssessingNeedsQuantum2022,hasanovicQuantumTechnicianSkills2022,masiowskiQuantumComputingFunding2022, greinertFutureQuantumWorkforce2023}, including our analysis of 34 interviews with industry representatives~\cite{greinertAdvancingQuantumTechnology2024}. This research provided the primary input for the extension of the \cf\ with the proficiency triangle and qualification profiles. The proficiency levels organized in the proficiency triangle cover key knowledge and skills discussed in the interviews, and the qualification profiles are supplemented by example personas/job roles and needs and suggestions extracted from the interview analysis.

The \cf\ provides a common language for QT-related qualification and education. It is a tool for planning educational activities such as workshops, training or study programs. It can be used to specify the qualification of an individual, or what is required for a job. For example, it is useful for ...
\begin{itemize}[leftmargin=1cm]
\renewcommand*\labelitemi{...}
    \item educators to plan educational activities and map their objectives;
    \item individuals to specify what qualification they have and what they want to achieve;
    \item learners to identify what training or course they should take;
    \item companies to specify what qualifications an employee or team covers and compare that to what they would need to identify what qualifications are missing;
    \item companies to specify job requirements and compare them with a candidate's qualifications;
    \item job seekers to map their qualifications and compare them to job requirements.
\end{itemize}
\vspace{0.5\baselineskip} 
Early users are educators~\cite[e.g.][]{hellsternIntroducingQuantumInformation2024,goorneyFrameworkCurriculumTransformation2024, patinoQuantumComputingEducation2024}. Two EU-funded projects agreed to use the framework to map their activities: DigiQ~\cite{shersonDigiQDigitallyEnhanced2024}, developing 16~QT master's programs across Europe, and QTIndu~\cite{qurecaQTInduQuantumTechnologies2024}, developing industrial training courses on QT. Additionally, first companies showed interest in using the \cf\ for their workforce development.

The \cf\ is an important step towards the standardization of QT education. This need was also addressed in one of the interviews analyzed for the update to version~2.5:
\vspace{0.5\baselineskip}

\begingroup 
\noindent
\leftskip 20pt \rightskip 20pt 
\small
\textit{I think it is always better to have a standard. So the question is how to standardize these trainings?
[...] It comes back to the ECTS systems from education. I mean, they are standards. %I think there should be an audit from the authority that comes and says `indeed, it is a good training'. 
}\hfill
~\cite[quote from an interview with a QT start-up;][p.~24]{greinertAdvancingQuantumTechnology2024}

\endgroup %empty line before \endgroup important
\vspace{0.5\baselineskip}
\noindent
For the update to version~2.0, we already aligned the proficiency level descriptions with the EQF and thus the European Credit Transfer and Accumulation System (ECTS), as described in Ref.~\cite[p.~3]{greinertQuantumReadyWorkforce2023a}. Following the EQF~\cite[p.~16]{europeancommissionEuropeanQualificationsFramework2018}, the three highest EQF levels are associated with the first, second and third cycles of the Qualifications Framework of the European Higher Education Area, which are based on the ECTS. 
Therefore, as the \cf\ levels are formulated based on the EQF, the level~B2 is linked to bachelor, C1~to master and C2 to doctoral programs. However, it should be emphasized that this mapping is not necessarily one-to-one, and that a degree program is not the only way to reach these levels. Especially in the business-related proficiency area (III), the high levels are more likely to be achieved through years of experience.

In version~2.0 of the \cf, these links were explicitly stated in the proficiency level descriptions. For example, it was noted for level B2: ``e.g. through a short research project as for a bachelor thesis, 
internship with project''~\cite[p.~5]{greinertEuropeanCompetenceFramework2023}. As mentioned above, this brings the risk of assuming a study program as the only way to reach a level, while there is a variety of opportunities to gain such a qualification. 

In version~2.5, such comments are less prominent and vary across the three proficiency areas. For the same proficiency level B2~Investigation, these are for area~(I) ``Short research project with documentation, e.g. student research project or bachelor thesis'', for area~(II) ``Write report on requirements analysis and performance results'' and for area~(III) ``Document an analysis for a concrete potential use case covering potential advances, risks, [...]%competative analysis, ...
''~\cite[p.~15-17]{greinertEuropeanCompetenceFramework2024a}.
With the update, the focus in these remarks shifted from an approximate duration to what an individual should be able to answer, perform or prepare for a qualification covering the corresponding proficiency level. Suggestions for training formats, the need for study programs or work experience are provided for the qualification profiles, rather than being fixed for a proficiency level.

The updated \cf\ is the starting point for a certification scheme to ensure comparability of industry training across Europe. This scheme is currently being developed within the Quantum Flagship coordination project, QUCATS~\cite{europeancommissionQuantumFlagshipCoordination2024}, associated to standardization activities. As in technology standardization, the first step is typically to address terminology, already at very low technology readiness, followed by test and measurement standards~\cite{deventerEuropeanStandardsQuantum2022}. Similarly, with the \cf\ we started to focus on the common language for QT education, based on the structured topics in the content map as well as the proficiency level descriptions. In version~2.5, there are already short, additional remarks on the aspect of testing or measuring if someone has reached a certain level. This aspect will be further elaborated in the certification scheme, a second step towards the standardization of QT education and qualification.

As a part of QUCATS, the EQRC highlights ``accords'' (best practices) of organizations across the EU~\cite[subpage Accords]{quantumflagshipEuropeanQuantumReadiness}. Within these accords, the \cf\ is a tool to provide  structure and standardization. As a recent example, the accord submitted by the QuantUM Group at the University of Minho provides a  profile of a software engineer in quantum computing as a representative graduate from the new master degree established there~\cite[subpage Accords: Educational Accords]{quantumflagshipEuropeanQuantumReadiness}.
In addition, the EQRC website hosts a curated playlist of videos, sorted based on the content map of the \cf~\cite[subpage Resources]{quantumflagshipEuropeanQuantumReadiness}.

The \cf, consisting of the content map and the proficiency triangle with associated qualification profiles, provides a reference framework for greater standardization across Europe. However, adoption is critical to its success.  Industry representatives seeking new employees should consider using the \cf\ to understand their needs and to formulate the 
required QT qualifications in their job advertisements. Educational institutions also have an important role to play. Here, the qualification profiles can be a step towards the future of standardized learning outcomes in QTs. 

Further research is needed on the application of the \cf. Whether it is used to prepare educational activities or within workforce development, accompanying research will be needed to identify issues and approaches for improving the \cf. In addition, regular updates will be needed to reflect technological changes and innovations as well as new challenges and needs in the quantum industry.

\begin{appendices}

\section{Structure and version history}
\subsection{Principal structure of the \cf}\label{app:structure}
\begin{figure}[hbt]
    \centering
    \includegraphics[width=0.8\linewidth]{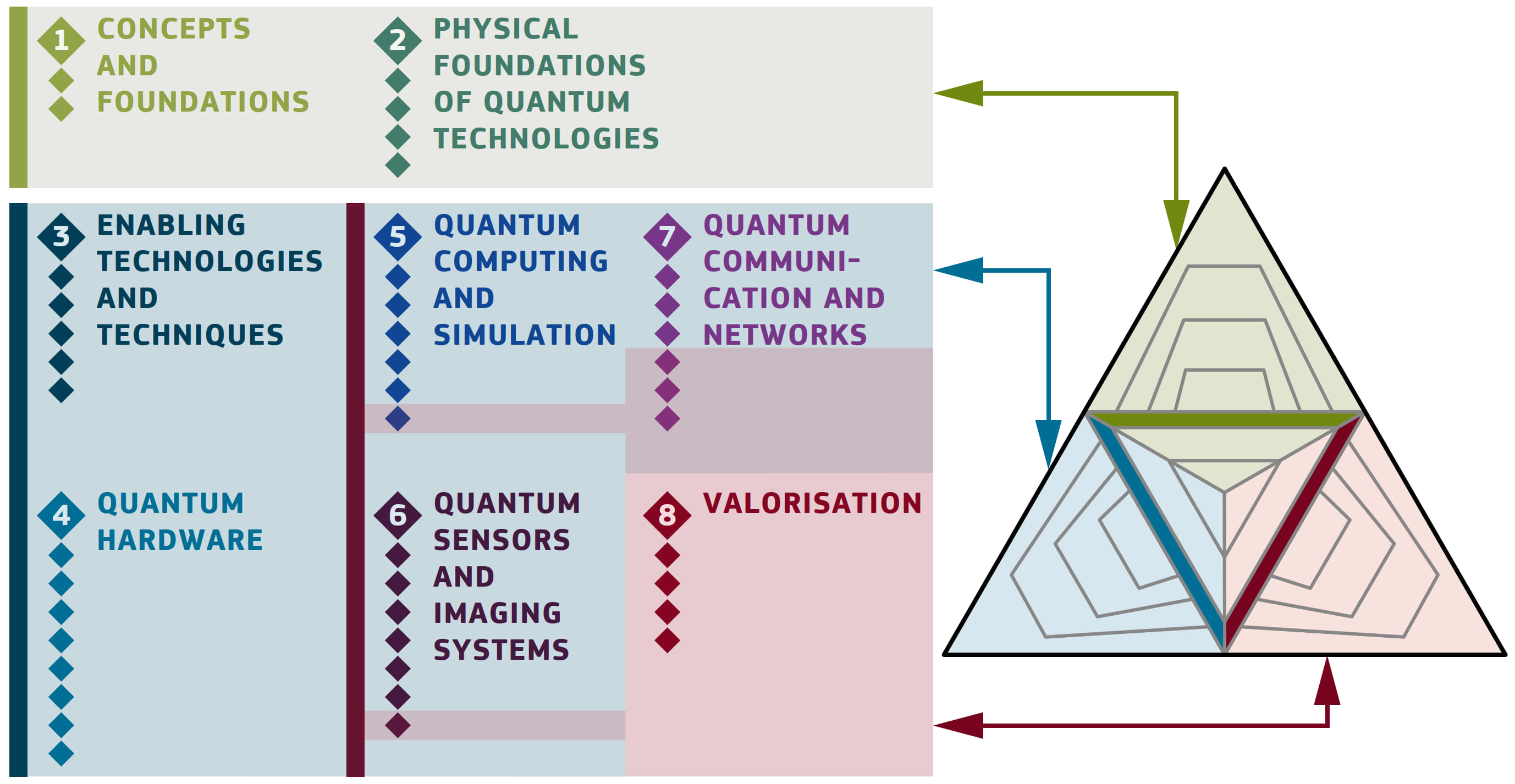}
    \caption[Coloring relation between content map and proficiency triangle.]{Coloring relation between content map (left) and proficiency triangle (right)~\cite[p.~4]{greinertEuropeanCompetenceFramework2024a}.}
    \phantomsection\label{fig:color}
\end{figure}

The \cf~\cite{greinertEuropeanCompetenceFramework2024a} consists of three main parts:
\begin{enumerate}
    \item The \textbf{content map} provides a structured overview of QT related topics and contents. It is a kind of extended table of contents with up to four layers: domains~1, 2,~...,~8 with 42~subdomains 1.1, 1.2, ..., (p.~5 in the \cf\ document) and, on the detail pages for each domain (p.~6–13), additional topics and sometimes subtopics provide more details for the subdomains. A selection of domains from the content map and of subdomains with topics are included in an example shown in Fig.~\ref{fig:QPexample}.  
    \item The \textbf{proficiency triangle} visualizes six proficiency levels for three proficiency areas (p.~14;  Fig.~\ref{fig:profTriangleEmpty}). In-depth level descriptions and more are provided for each of the three proficiency areas (p.~15-17). See Sec.~\ref{sec:PT}. 
    \item The \textbf{qualification profiles} show typical qualification, i.e. coverage of the proficiency triangle. An overview (p.~18) and detailed descriptions including example personas and (training) suggestions (p.~19–27) are available. See Sec.~\ref{sec:QP}.
\end{enumerate}
Furthermore, two examples how to combine a profile with a content selection are given (p.~28–29 in the \cf). In addition, the \cf\ pages 3 and 4 provide some information on how to use the framework, including an introduction of the terminology and the level system, and information on the methodology and related publications.
The coloring of the proficiency triangle corresponds to the coloring of the content map as shown in Fig.~\ref{fig:color}.

\subsection{Version history and related publications}\label{app:versHis}
For the Qualification Profiles, a beta version was published in 2022~\cite{greinertQualificationProfilesQuantum2022} as a separate document based on the \cf\ version~1.0. The new profiles replace this beta version, they are not an update of the beta version and were created with a completely different approach. While four of the six profiles in the beta version were those of an engineer in QT hardware or one of the QT pillars, the new profiles focus on the proficiency that a QT engineer needs in general. These four beta version profiles would be the same profile in the new version. Now the pillar is the context to be added by a topical selection of the content map. In the beta version, the profiles consisted mainly of the topical selection of the content map, and only provided a very limited specification with proficiency levels, which themselves were not specified at that time.

Version~2.5 of the \cf\ was released in April 2024 on Zenodo~\cite{greinertEuropeanCompetenceFramework2024} and in August 2024 as an `official version' by the Publications Office of the European Union~\cite{greinertEuropeanCompetenceFramework2024a}. They differ only in the placement or addition of some editorial information (mainly moved from page 3 to the new page 1), but the framework itself has not changed between the EU and Zenodo publications.
Table~\ref{tab:versions} provides a version history and an overview about related publications.
    
\begin{table}[h]
\centering 
    \caption[Version history and publications related to the CFQT]{Version history and publications related to the \cf, 
    with \cf\ standing for `European Competence Framework for Quantum Technologies' and `QT(s)' for `Quantum Technology(-is)' (in the titles). [*] Older versions of the \cf\ are also available on Zenodo~\cite{greinertEuropeanCompetenceFramework2024}. \\
    }\phantomsection\label{tab:versions}
    \begin{tabular}{p{0.15\textwidth-2\tabcolsep}
			p{0.35\textwidth-2\tabcolsep}
			p{0.5\textwidth-2\tabcolsep}}
        Date & [Ref.] Title/version & Comment \\
        \hline
        Dec. 2020  & \cite{greinertneegerkeBetaVersionEuropean2020} Beta version of the \cf\ & Content map with seven domains and details pages, one keyword for each of six proficiency levels. Available on request on Zenodo.  \\
        June 2022 \newline (Nov. 2020) & \cite{gerkeRequirementsFutureQuantum2022} Requirements for future quantum workforce – a Delphi study & Interim report on the iterative study (Ref.~\cite{greinertFutureQuantumWorkforce2023}), conference 
        proceedings of GIREP, 
        Nov.~2020. \\
        May 2021 & \cf\ version 1.0 (before graphical update)& Update of content map, restructuring the QT pillar domains to a total of eight domains. 
        \\
        Sep. 2021 & [*] \cf\ version 1.0 & Only graphical update.
        \\
        Sep. 2021  & \cite{greinertCompetenceFrameworkQuantum2021a} Competence framework for quantum technologies: meth\-od\-ol\-o\-gy and version history & Documentation of the development process until version~1.0, published by the EU Publications Office.
        \\
        Jan. 2022  & \cite{greinertQualificationProfilesQuantum2022} Qualification Profiles for QTs (beta version) & Six profiles: content map selection with proficiency level indication, 
        outdated.
        \\
        June 2023 \newline (Aug. 2022) & \cite{greinertFutureQuantumWorkforce2023} Future quantum workforce: Competences, requirements, and forecasts & Paper on the iterative study collecting input for the \cf\ beta version and also version~1.0, submitted in Aug. 2022\\
        April 2023  & [*] \cf\ version 2.0 & Update of content map, addition of descriptions for six proficiency levels. 
        \\
        July 2023  & \cite{greinertQuantumReadyWorkforce2023a} Towards a quantum ready workforce: the updated \cf & Paper on the update to version~2.0. \\
        April 2024 & \cite{greinertEuropeanCompetenceFramework2024} \cf\ version 2.5 (Zenodo)& Extension by proficiency triangle, new qualification profiles, content map unchanged. 
        \\
        July 2024 & \cite{greinertAdvancingQuantumTechnology2024} Advancing QT workforce: industry insights into qualification and training needs & Preprint on the analysis of interviews with industry that were also used to update the \cf\ to version~2.5\\
        Aug. 2024 & \cite{greinertEuropeanCompetenceFramework2024a} \cf\ version 2.5 & Official published by EU publications office.
        \\
    \end{tabular}
\end{table}
%\clearpage
\section{From competence types through categories to qualification profiles}

\subsection{Competence types}\label{app:material_CT}
\begin{figure}[htp]
    \centering
    \includegraphics[angle=90,
    width=0.67\textwidth]{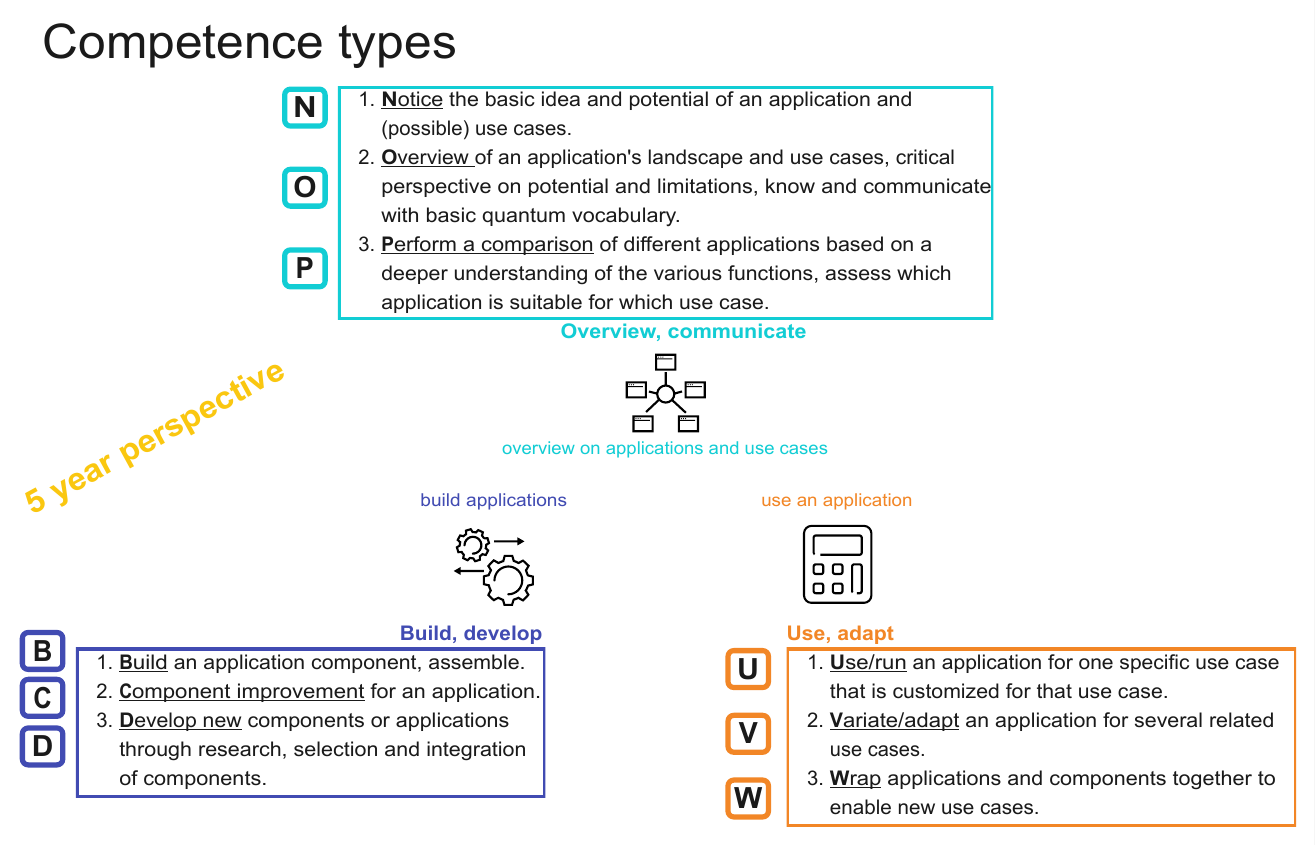}
    \caption[Interview material `competence types'.]{Competence types as a material shown in the interviews, June 2023, outdated. For better readability, the descriptions are reproduced in Sec.~\ref{subsec:methods:compTypes}.}
    \phantomsection\label{fig:CompTypes}
\end{figure}

In some of the interviews (22 out of 34), the slide/material `competence types' was shown, reproduced in Fig.~\ref{fig:CompTypes}. They formed the initial system that was refined and extended to the proficiency triangle through the \cf\ update process. %, as described in Sec.~\ref{subsec:methods:compTypes}.

\subsection{Categories and qualification profiles}
\label{app:categories_QP}

\noindent
As described in Ref.~\cite{greinertAdvancingQuantumTechnology2024}, the interviews  34 interviews with industry representatives were analyzed using qualitative content analysis.  Table~\ref{tab:categories} shows an extract from the category system related to the qualification profiles (left column) as well as a revised version of these categories used for a second iteration of the categorization (middle column). In addition, the qualification profiles in the \cf\ version~2.5 are listed (right column). 
For the first iteration, also the related `competence type' level is given, e.g. [B] for `Build an application...', as specified in Fig.~\ref{fig:CompTypes}/Sec.~\ref{subsec:methods:compTypes}. 
The progression from competence type levels to qualification profiles is visible in rows.
\vspace{-0.1cm}
\newcommand{\Einzug}{\quad \hangindent 1.0em}
\newcommand{\EinzugZwei}{\quad \quad \hangindent 2.0em}
\begin{table}[ht]
    \centering
    \caption[Category system and statistics regarding the qualification profiles.]{Category system and statistics for the two iterations related to the development of the qualification profiles together with the final profile number (No.) title in the \cf\ version~2.5. In the columns called N, the number of segments categorized in the category is given. If no number is given, the code is a supportive code only to structure the codes on the next level. %With `incl.' for including, `comp.' for `company', `eng.' for `engineer', `q.' for `quantum'.
    }
    \begin{tabular}{p{4.1cm}@{\hskip 6pt}c|p{3.73cm}@{\hskip 6pt}c|c @{\hskip 6pt}p{2.2cm}}
         First iteration &&Second iteration &&\multicolumn{2}{l}{ Qualification Profile}\\
        Category & N &Category & N & No. &Title in CFQT\\
        \hline
        %general or content-focused remarks, challenges &6&general or content-focussed remarks, challenges &9&&\\ \hline
        quantum aware workforce && &&&\\
            \Einzug basic user (click a button) 
            [U]% -- use]
            &6&\multirow{2}{3.73cm}{other: not quantum specific/ no quantum needed} &16&&\\
            \Einzug engineers with (almost) no quantum 
            [B]% -- build]
            &22&&&&\\
             \Einzug hype, basic idea (not) for all, incl. admin people 
            [N] % -- notice]
            &24
            &QT aware person &18&P1&QT aware person\\
            & &QT aware decision maker, manager &17&P2&QT informed decision maker\\
        \hline
        quantum literate workforce && &&&\\
            \Einzug management (non-QT comp.), business, sales, policy makers
            [N/O]% -- notice/overview]
            &42&QT literate (sales) communicator &27&P3&QT literate person\\
            \Einzug use-case identifier
            [O/W]
            &19&\Einzug use case identification&5&&\\
            \Einzug eng. with overview on QT (also e.g. cryo physicists; incl. programming)
            [O/B]
            &73&QT user, engineer working with QT &69&P4&QT practitioner\\
            \Einzug advanced user (with adaption)
            [V]
            &23&\Einzug QT user (with adaptation)&19&&\\
        \hline
        q. expert workforce (needs QT study program or PhD) && &&&\\
            \Einzug strategist, consultants 
            [P] (comparison, assessment)
            &27&QT analyst (market, use cases, strategies) &15&P5&QT business analyst\\
             &&QT engineer (generalist) &6&P6&QT engineering professional\\
         \hline
             &&senior QT engineer, QT architect &&P7&QT (HW or SW) specialist\\
            \Einzug customer/sales eng.
            [C/V]
            &12&\Einzug ... focus on customer and applications (incl. consulting, perform comparison)&21&&\\
            \Einzug quantum algorithm developers (high-level)
            [D]
            &22&\EinzugZwei QT application specialist (e.g. programming)&22&&\\
            \Einzug systems eng. (integration, industrialization)
            [C]
            &13&\Einzug ... focus on systems and integration
            &12&&\\
             &&QT strategist (e.g. consultant) &2&P8&QT (product) strategist\\
            \Einzug quantum experts/\newline expertise needs
            [D]
            &35&QT core innovator &29&P9&QT core innovator\\
    \end{tabular}
    \phantomsection\label{tab:categories}
\end{table}

\subsection{GPT extracted job roles}\label{app:GPTpersonas}
As described in Sec.~\ref{subsec:methods:jobRoles_QualiProf}, 30~anonymized interview transcripts were provided to an LLM (gpt3.5-turbo-16k) to extract ten personas as typical job roles. The prompts were:
\begin{description}
    \item[Prompt 1:] \textit{Here is an anonymised interview transcript. What are the key personas described in the interview? (e.g Engineers, Developers, Project Managers)?}\newline
    The outputs of prompt 1, for each transcript, were collated into a single list.
    \item[Prompt 2:] \textit{This document contains the key personas described in a series of interviews with companies in the Quantum Technology industry. Each interview begins with the text ``INAXXX''. Based on these findings, please provide a list of 10 types of personas in the Quantum Technology industry, each with a 2 sentence description.}
\end{description}

\noindent
The extracted roles are:
    \begin{enumerate}
        \item Quantum Engineers/Researchers/Scientists: These individuals are involved in the design and development of quantum technologies, requiring expertise in quantum physics, engineering, and material science.
        \item Project Managers: Responsible for overseeing the successful development and implementation of quantum-enabled devices, coordinating efforts between different teams or departments.
        \item Decision Makers/Management: Individuals in managerial or executive roles responsible for making strategic decisions based on market needs and technology trends in the quantum industry.
        \item Training and Development Professionals: Facilitate the learning process for existing employees or new hires, designing hybrid training programs, online content, workshops, or lectures.
        \item Generalists: Individuals with a broader understanding of quantum technologies, bridging the gap between technical and market-oriented aspects of quantum technologies.
        \item Market Analysts: Assess market demand and guide the development of quantum-enabled devices by understanding market trends, potential applications, and customer needs.
        \item Company Representatives: Engaged in technology development, consulting, and strategic planning for quantum technology companies, addressing workforce development and skilled personnel needs.
        \item End-users: Users in various industries, such as medical imaging, interested in understanding and benefiting from quantum-enhanced technologies.
        \item Technical Developers: Involved in building physical devices and components for quantum technologies, such as lasers and high-precision optical and frequency components.
        \item Sales Managers and Marketing Personnel: Responsible for understanding the market landscape, guiding sales engineers, shaping product strategy, and communicating the value of quantum technologies to potential customers.
    \end{enumerate}
These personas were used to refine the initial categories from the qualitative content analysis of the interviews (see Table~\ref{tab:categories}) and thus influenced the preparation of version~2.1 of the qualification profile descriptions. 

\end{appendices}

\section*{Abbreviations}
\begin{itemize}[leftmargin=2cm]
    \item[CFQT] European Competence Framework for Quantum Technologies~\cite{greinertEuropeanCompetenceFramework2024a}
    \item[CSA] Coordination and Support Action
    \item[DigiQ] Digitally enhanced quantum technology master (project title) \cite{shersonDigiQDigitallyEnhanced2024}
    \item[EQF] European Qualifications Framework~\cite{europeancommissionEuropeanQualificationsFramework2018}
    \item[EQRC] European Quantum Readiness Center \cite{quantumflagshipEuropeanQuantumReadiness}
%    \item[GPT] Generative Pre-trained Transformer (one type of LLMs)
    \item[HW/SW] Hardware/Software
    \item[LLM] Large Language Model, e.g., GPT (Generative Pre-trained Transformer)
 %   \item[ML] Machine Learning 
%    \item[QKD] Quantum Key Distribution 
    \item[QT/QTs] Quantum Technology/-ies
    \item[QTEdu] Quantum Technology Education (QTEdu CSA, project title) \cite{qteducsaQuantumTechnologyEducation}
    \item[QTIndu] Quantum Technologies courses for Industry (project title) \cite{qurecaQTInduQuantumTechnologies2024}
    \item[QUCATS] Quantum Flagship Coordination AcTion and Support (project title)~\cite{qucatsQuantumFlagshipFuture2024}
    \item[STEM] Science, Technology, Engineering, and Mathematics
    %\item[SW] Software
\end{itemize}

\backmatter

\section*{Declarations}
\bmhead{Data availability}
The three versions v2.1-v2.3 of profile descriptions %, the 50 job posts, the prompt and the mapping statistics 
are available on Zenodo~\cite{greinertSupplementaryMaterialAdvancing2024}. Here, also data from the industry needs analysis (Ref.~\cite{greinertAdvancingQuantumTechnology2024}) is available.

\bmhead{Acknowledgements}
We thank Carolyn Levy, Dennis Green, Dion Timmermann, and all experts from industry and academia for their valuable feedback within the iterative refinement of the descriptions of the proficiency level or qualification profiles. %, as well as all others that provided feedback. \newline
%The \cf\ is intended to be a living document with regular updates, feedback is welcome at any time. 

\bmhead{Author contributions}
FG prepared, conducted and analyzed the interviews, prepared the \cf\ drafts, collected feedback and wrote the original manuscript. 
SG conducted the LLM-based analysis (Sec.~\ref{subsec:methods:jobRoles_QualiProf}). % and wrote the related part of the original manuscript (Sec.~\ref{subsec:Simon} and \ref{sec:ResFromMapping}). 
All authors contributed to the study design and analysis, i.e. they discussed the methods and results, revised the manuscript, and read and approved the final manuscript.

\begin{wrapfigure}[3]{r}{1.5cm}  
\vspace{-4ex}
    \includegraphics[width=1.5cm]{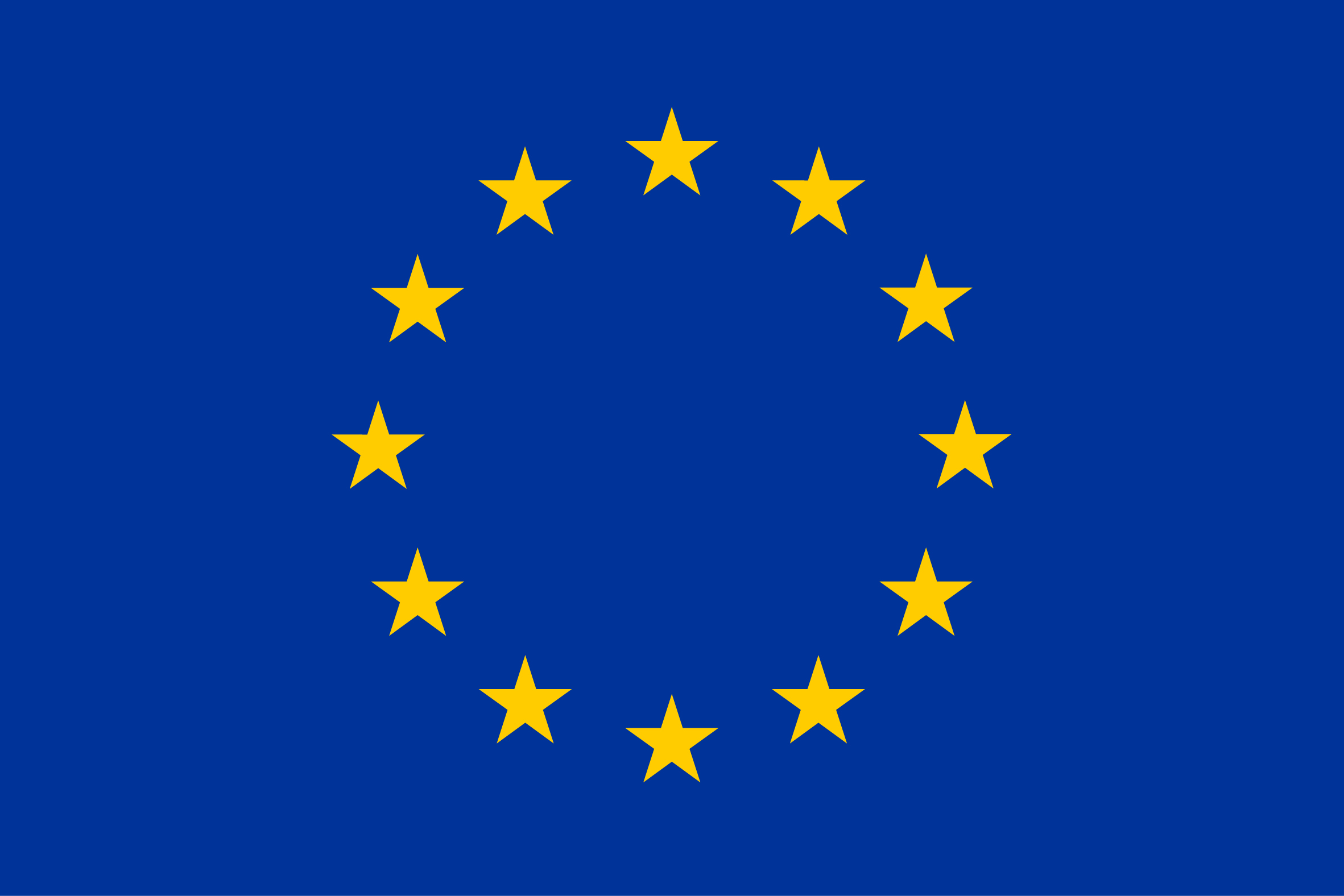}
\end{wrapfigure}
\bmhead{Funding}
This work has received funding from the European Union’s Horizon Europe research and innovation programme under grant agreement No 101070193.

\noindent This publication reflects only the views of the authors, the European Commission is not responsible for any use that may be made of the information it contains.

\bmhead{Competing interests}
The authors have no competing interests to declare that are relevant to the content of this article.

\bmhead{Financial Interests}
The authors have no relevant financial or non-financial interests to disclose.

\bmhead{Ethics approval}
Ethical review and approval was not required for the study on human participants in accordance with the local legislation and institutional requirements.

\bibliography{sn-bibliography}% common bib file

%\clearpage
%\newpage


\end{document}